# Peer-to-Peer Network Simulators: an Analytical Review

Mansoor Ebrahim, Shujaat Khan and Syed Sheraz Ul Hasan Mohani

*Abstract*— Simulators are the most dominant and eminent tool for analyzing and investigating different type of networks. The simulations can be executed with less cost as compared to large-scale experiment as less computational resources are required and if the simulation model is carefully designed then it can be more practical than any well brought-up mathematical model. Generally P2P research is based on the principle of simulate first and then experiment in the real world and there is no reason that simulation results cannot be reproducible. A lack of standard documentation makes verification of results harder as well as due to such poor documentation implementation of well-known overlay algorithms was very difficult. This Paper describes different types of existing P2P simulators as well as provides a survey and comparison of existing P2Psimulators and extracting the best simulator among them.

## I. INTRODUCTION

P2P network has revolutionised the concept of client server model [1] by allowing users to act as both clients and servers and by facilitating scalability, usability, efficiency, and performance as major features. P2P networking is a utility application that runs on a personal computer and share files with other users across the web. P2P is a wide and significant field for research and is the expected consequence of the Internet and business-related efforts to interconnect computer systems for competent sharing of information.

## II. P2P SIMULATOR

In this section, different simulators have been analyze on the basis of certain parameters which includes Simulator Architecture, Usability/ Documentation, Scalability, Statistics, portability, and System Limitations and by analyzing these parameters, the best among them is selected.

Mansoor Ebrahim is with the Faculty of Engineering, Sciences & Technology, Iqra University, Karachi, Pakistan. Email: mebrahim@iqra.edu.pk
Shujaat Khan is with Faculty of Engineering, Sciences & Technology, Iqra University, Karachi, PakistanEmail: shujaat123@gmail.com
Syed Sheraz Ul Hasan Mohani is with the Faculty of Engineering, Sciences & Technology, Iqra University, Karachi, Pakistan. Email: smohani@iqra.edu.pk

### A. Criteria

The criteria on which the different simulators are being analysed are;

**Simulator Architecture** defines the structure and operations that the simulator can perform, its characteristics and how they are implemented. It also determines that the simulator has either a discrete event simulation engine or query-cycle that is whether it makes use of a scheduler which synchronizes the messages shared among nodes or the simulation loops through each node, carrying out queries for the nodes as required.

**Usability/Documentation** One of the key parameter in examining of simulator is to define that whether the simulator is user friendly or not i.e. how simple is to use and learn the simulator, this includes that the simulator source code should be well commented and should execute a clean, well designed and documented API. Manuals, user guides and other appropriate documentation should also be provided. Online support such as mailing lists, newsgroups and websites should be available to facilitate user. These criteria also includes the way in which experiment set-ups are created, if there is a script language, what type of documentation is provided and how simple it is to follow [3][4].

**Scalability** As P2P protocols are designed in such a way that they provides scalability and are able to solve problems related to scalability. Scalability is one of the major parameter on which a simulator can be analyzed.

**Statistics** Another major parameter in analyzing the simulators is the output it generates. The output needs to be closed to the desired assumptions and simple to operate so that statistical analysis can be carried out and to produce graphs. The simulator should be able to provide snapshots of its state for analysis. The simulation script file should allow for reproducible experiments [3][4].

**Portability** It should be possible to reuse simulation code with minor modification [3] [4].

**System Limitations** defines how fine the simulator works by make use of the computer resources available to it. If it doesn't use the resources efficiently then its ability to scale is reduced [3] [4].

### B. Evaluation Methodology

The evaluation methodology was easy and simple. The latest version of each simulator was downloaded as well as resources such as manuals, source code and research papers were studied, each simulator was evaluated on the basis of above-mentioned parameters. Not a single simulator fully contented the evaluation criteria, with some having greater deficiencies than others. Of most concern was the lack of support for collecting statistics from simulation runs. Authors of different simulators claims to the scalability of their simulators, tests were carried out on the scalability of some of the simulators by performing experiments in order to find the maximum number of nodes that can be linked to the network and interact successfully. But even if achievable, these are significantly less than the numbers that can be achieved in real P2P systems [3-4].

*C. Simulator Review*

*1) PeerSim:* PeerSim an event-based P2P simulator, partly developed under the BISON project and released under the GPL open source license. It is written in Java and exclusively designed for epidemic protocols. It offers both cycle-driven simulation and discrete-event simulation using separate simulation engines. It has been designed to be scalable and dynamic for simulating large P2P networks. Structured and unstructured overlays can also be simulated by using PeerSim[5-6].

**Simulator Architecture**

PeerSim supports two models of simulation: Cycle based and Event-based models, which can be attained by using two of its different simulation engines: a cycle-based engine and an event-based engine. Cycle-based model is based on random selection of nodes and each node protocol is call upon in turn at each cycle. Whereas event-based model is based on scheduling set of messages in time and node protocols are call upon according to the time message delivery order [5]. It uses command based environment and inputs are fixed with no changes on run time.

**Statistics**

PeerSim provides neither a graphical user interface nor any debugging facilities. However, it offers class packages to perform common statistics calculation as well as additional user-defined data collection coding. PeerSim offers some class packages that support some renowned models such as lattice and random graph. It also has a class package that performs statistical computations [4-5].

**Scalability**

PeerSim can simulate millions of network nodes, and it is clamed that PeerSim can hold up a network up to 106 nodes in the cycle-based model [5].

**Usability /Documentation**

The PeerSim website offer some tutorials and API documentation. However, The C++ API documentation is poor; the most of the tutorials and examples are based on the cycle-based model with no in-depth discussion on the event-based model. It is not straightforward to utilize this model [4-5].

**Portability**

PeerSim allows user-defined entities to replace almost all predefined entities in it. It bears extendable and pluggable component features. Plain ASCII file is used for flexible configuration comprising of key-value pairs [5].

**System Limitations**

No support for distributed simulation. Node identifications can be modified by user-defined mechanisms and are produce as integers [5].

**Analysis**

PeerSim is command based. The inputs to the simulator are predefined (set by default) in a text file, so user cannot change the inputs on run-time. As there is no GUI it is very hard to use and the generated results are difficult to analyze. The API documentation is good but C++ API documentation is poor as it only defines the basics in terms of how to compile and run the simulator. There is no in-depth discussion regarding the extension of the simulator's source code. The results generated by PeerSim are not sufficient enough to perform good statistical analysis.

*2) GPS:* GPS is a message-level event driven P2P simulation framework aimed at modelling P2P protocols accurately and efficiently in a realistic and dynamic environment. For the purpose of portability, ease of development and extensibility, GPS uses Java as the development and simulation language [4].

**Simulator Architecture**

GPS is a discrete-event, rather than time-driven, simulator. Instead of advancing the simulation time in fixed increments and processing events synchronously at each clock tick, processing and time advancement is triggered by the occurrence of events. GPS maintains efficiency by modelling communication at the message level. On the other hand, accuracy is maintained by taking into account the underlying network and protocol properties (in this case TCP) without the overhead of packet level simulation; this is achieved by the use of macroscopic models of estimating performance [8].

**Statistics**

GPS is extensible for modelling any P2P protocol, integration with GUI and network visualization and provides support for topology generation tools. It provides simulation models for Bit Torrent, which, to the best of our knowledge has not been modelled functionally before [3].

**Scalability**

GPS is a message level simulator that also models the underlying network topology using the GT-ITM model. As it models the network topology partly so each packet is not modelled, however it supports different number of flow level models [5].

**Usability /Documentation**

The API is poorly documented and as there is no in-depth discussion it is very difficult to run the simulator. It is not straightforward to utilize this mode [5].

**Portability**

For the purpose of portability, ease of development and extensibility GPS uses Java as the development and simulation language. To support accurate simulation GPS models the network topology and characteristics. The GPS framework provides all the infrastructures required for P2P simulation, so new protocols can be easily plugged in and even run on existing protocols [8].

**System Limitations**

Limited Support for extending the simulator for protocols other than Bit [2] Difficult to implement any other protocol as the simulator is tightly coupled to the Bit Torrent protocol [3, 5]

**Analysis**

GPS provides a GUI environment to interact with the users. GPS has the capability to allow users to select macro models from a list of models. The API documentation of the GPS doesn't contain sufficient information about the simulator and as there is no in-depth discussion it is very difficult to run the simulator, hence the extensibility of the code is not easy. Since there isn't any user manual, the simulations do not provide enough information to determine the relation between the inputs and the corresponding generated outputs. Hence, it is very difficult to do analysis on the simulation results. Another drawback of GPS is that it only supports BitTorrent protocol and not any other protocol.

*3) NeuroGrid:* The NeuroGrid simulator was originally developed as a simulator for comparing the performance of the Freenet, Gnutella and NeuroGrid protocols. More recently, the simulator has been extended to support DHT protocols such as Pastry. Indeed, the NeuroGrid simulator was designed so as to be as extensible as possible with regards to supporting new protocols [2, 7].

**Simulator Architecture**

Neurogrid is a single-threaded discrete event simulator, initially deliberated for protocol comparison. The simulator operates on the overlay layer level and can simulate either structured or unstructured protocols [5]

**Statistics**

It provides flexible means for gathering statistics, by allocating comprehensive data to be extracted. Simple file parameters are used to specify simulation scenarios. Generate simulations for pre-determined variables, and code would have to be modified for others. Only the overlay layer can be simulated with this simulator. It assumes a graph topology as input to the simulator [5, 7].

**Scalability**

One of the simplifying assumptions made by the NeuroGrid simulator is that all links between nodes have equal bandwidth and no bandwidth data is associated with the list of connected nodes stored at each node. This simplifying assumption allows simulating up 300,000 nodes on machine having almost 4GB RAM, but replication fails due to thread limits [5-6].

**Usability /Documentation**

Documentation is extensive on the web but a little disorganized in wiki form.

**Portability**

NeuroGrid simulator was mainly created with extensibility in mind. The Simulator makes use of a number of abstract classes that are intended to be generic across different P2P implementations. [9].

**System Limitations**

Simple file parameter file does not include the capability to schedule events at particular times. It may be simple to amend the simulator to implement behaviour, but Node failure is not at present implemented.

**Analysis**

Initially NeuroGrid seems to be a better option for our P2P simulations as the documentation for NeuroGrid system was extensive to understand the working of the simulator.

*4) Query Cycle:* The Query Cycle Simulator (sometimes called P2PSim) is a peer-to-peer simulator developed by Stanford University. The primary focus of the Query Cycle Simulator is to accurately model user behaviour in a peer-to-peer file-sharing network [3].

**Simulator Architecture**

The Query-Cycle is a query-based simulator written in java, specialized in file-sharing simulations. It includes realistic models for content distribution (for both data and types), query activity; download behaviour, uptime, etc. The simulation engine is based on the following model. The simulation proceeds in query cycles. In each query cycle, a peer may be actively issuing a query, inactive, or even down and not responding to queries passing by. Upon issuing a query, a peer waits for incoming responses, selects a download source among those nodes that responded and starts downloading the file. The query cycle finishes when all peers who have issued queries download a satisfactory response. Statistics may be collected at each peer, such as the number of downloads and uploads of the peer [11].

**Statistics**

Query-Cycle provides a graphical user interface in which user can set different parameters for the network description, content allocation and the peer behaviours. Once the parameters are set and the simulation is started, the properties cannot be customized; however there is possibility of restart, halt, and save the simulation. There is also a visualize for presenting the state of the network. Statistics computation involves the number of uploads and downloads at each node [5].

**Scalability**

Query cycle simulator can simulate millions of network nodes and its performance is exemplary in terms of modelling of peer behaviour, however it suffers from limited scalability. According to one of the authors QCC does not scale well above 1000 peers, although this is apparently due to the fact that QCC "models actual [3].

**Usability /Documentation**

The Query Cycle developer website provides demo tutorial of the software and API documentation.

**Portability**

Query-cycle allows user-defined entities to replace almost all predefined entities in it. It bears extendable and pluggable component features

**System Limitations**

No specific system limitation is defined.

**Analysis**

Query Cycle Simulator provides a GUI interface for a user to run simulations. Since the user interaction is GUI based, so the usability of is not difficult in terms of interoperability of the simulator. The documentation contains API documentation only and there is very few information on the user manual and also there isn't any sufficient information on how to compile the simulator code. It seems that the source code is extendible but lack of proper documentation does not support its extensibility. Statistical analysis is only limited to the simulator's predefined analysis i.e. the simulator does not allow the use of generated results to be used for further analysis using a different tool. One of the major analyses done by the simulator is the rate at which number of downloads/uploads occurs at each peer.

*5) P2PSim:* The primary goals of P2PSim are "to make understanding peer-to-peer protocol source code easy, to make comparing different protocols convenient, and to have reasonable performance". P2PSim is one of the few peer-to-peer simulators to make use of threads in simulation. Gil et al. state that the use of threads makes "the implementations [of peer-to- peer protocols] look like algorithm pseudo-code, which makes them easy to comprehend". P2PSim uses an event queue to store pending events sorted by time stamp [12].

**Simulator Architecture**

P2PSim is a discrete event simulator that is used to simulate structured overlays only.. P2Psim can simulate node failures and both iterative and recursive lookups are supported. 160-bit SHA-1 hashing produces node IDs. No support for Distributed simulation, cross traffic and enormous fluctuations of bandwidth [5].

**Statistics**

Limited set of statistics can be collected before coding required. The P2PSim provides an extensive variety of underlying network topologies such as random graph, end-to-end time graph, G2 graph, and Euclidean graph. Euclidean graph is the most commonly used [5].

**Scalability**

P2PSim scalability has been tested with a 3,000-node Euclidean constant failure model topology; moreover simulation on 1,700-node Internet topology has also been performed with huge data set [5].

**Usability /Documentation/ Portability**

The C++ API is poorly documented, but by extending certain base classes other protocols can be developed and implemented [5].

**System Limitations**

Lack of support for simulating unstructured or semi-structured P2P routing protocols [6].

*6) 3LS:* The 3LS or 3 Layered System peer-to-peer simulators was developed by N. Ting of the University of Saskatchewan. Ting developed a set of criteria for evaluating peer-to-peer simulators. These evaluation criteria are usability (ease of use), extensibility (supporting different protocols), configurability (setting configuration parameters), interoperability (with other applications), level of detail (of simulated environment) and build-ability (simulated code can be deployed as working application).

**Simulator Architecture**

3LS is an open-source simulator for overlay networks designed to overcome the problems of extensibility and usability. The system is separated to three architectural levels: a network model, a protocol model and a user model. The network model uses a two-dimensional matrix as storage of distances between the nodes. The protocol model defines the current protocol being simulated. The user model is the input interface for the user [3].

3LS' name stems from the idea of splitting the simulator into 3 separate levels; the network level is at the bottom, the protocol level is in the middle and the user level is on top. Communication can only occur between adjacent levels. The network level contains a two-dimensional matrix (i.e. adjacency matrix), which stores the distance values between the nodes in the simulation. Worker threads are associated with each of the nodes to perform communication with other nodes in the network. Four queues are used at each node to store pending message objects. These queues are the Outbox, the Inbox-for-network-delay, the Inbox-for- processor-delay and Inbox. The use of these queues is intended to allow the simulation of delays related to network traffic and CPU delays. A 'static step clock' is used to propagate messages between the various queues, and a node receives the message when it reaches its Inbox. The protocol level contains instances of a peer class. Any protocol implemented by the simulator must extend this class. The Gnutella peer class is used as an example, and apparently this object is used for maintaining the list of a peer's neighbouring peers. Each instance of the protocol class is also linked to an application class, which is responsible for passing message objects to other peers. The user level was not actually implemented, although the idea seems to be to provide a means for scheduling events related to user specific behaviour [12].

**Statistics**

3LS is an open source simulator, integration with GUI and uses main memory to store each event executed for visualization.

**Scalability**

The major drawback of 3LS is scalability: memory overhead incurred by the Network layer and the GUI (Graphical User Interface) is too high, which restricts the simulated network's size to a couple of thousand peers on a regular machine [3].

The 3LS uses most of the memory resources to a graphical interface as the simulator uses main memory to store each event executed for visualization and this limits the system to less than a thousand nodes on a low-cost workstation [3].

**Usability /Documentation**

The 3LS API are documented, but the simulator is not available on the web.

**Portability**

3LS is an open-source simulator for overlay networks designed to overcome the problems of extensibility and usability [3].

**System Limitations**

The simulator makes use of main memory to store visualization events, which in result limits the system performance in terms of nodes.

**Analysis**

The main drawback of this simulator is that it is currently in inactive state. The source code is only available via requests made to the authors of the simulator. All the above-mentioned information about this simulator is on theoretical basis as provided by the author. 3LS has been mentioned in the existing research papers on simulator survey, which shows that any P2P research group has not implemented it.

*6) RealPeer:* RealPeer is a development framework for P2P systems. Utilizing this framework a developer can build a P2P application (as the central element of a P2P system) that can be both: executed as a simulation model on a local computer interacting with thousands of other simulated peers on the same computer as well as can be executed as a real P2P application that connects to remote peers (for instance on the internet). Thus, a simulation model of a P2P system can be reused as part of a real P2P system and vice versa [13].

**Simulator Architecture**

The RealPeer framework has the following characteristics:

The framework supports the modelling and simulation as well as the development of P2P systems. It is as generic as feasible with regard to the class of P2P systems that can be modelled, simulated and developed using the framework. The framework offers one single representation for a P2P application, which permits that application to be reused both as a simulation model and as part of a real P2P system. The framework's architecture is highly modular and extensible with a clear separation of concerns. This enables a developer to combine and freely exchange elements of the framework and the model and thus provides a mechanism to reuse elements. In order to support the simulation of large-scale P2P systems, the framework is as scalable and lightweight as possible. The framework enables a developer to conduct controlled simulation experiments with complete internal validity in order to obtain accurate and reproducible simulation results [13].

**Scalability**

RealPeer developers have tested its scalability with a 20.000 peers. The simulator is sequential (parallel and distributed simulation may be incorporated in a future version).

**Statistics**

No GUI is available for real peer until now and the simulations are based on command bases.

**Usability /Documentation**

Documentation is available on the web.

**Portability**

Real peer is an object-oriented software framework with heavy use of the plug-in design pattern, which allows extensibility in the simulator

**Analysis**

RealPeer makes use of command-line program in order to run simulations. Although it also provides GUI based predefined simulations for some example scenarios. The simulations results are logged in a text file but those results do not prove to be meaningful as only numeric values are stored without any corresponding variable name. The inputs to the simulator are to be defined in text file that shows real time simulation is not supported. One of the positive aspects of this simulator is that it is relatively new and in active state.

*7) PeerThing:* PeerThing [15] simulator is an application for modelling and simulating peer-to-peer-networks. PeerThing is written on Java and a tool known as Eclipse is used to provide a GUI interface.

**Simulator Architecture**

This chapter describes the architecture of PeerThing simulator. The architecture is basically divided into two major categories: "system behaviour" and "system scenario". The system behaviour allows defining the behaviour of each peer in a P2P network and scenario allows defining the environment of the P2P network. The system behaviour takes a peer-centered approach that makes use of behaviour of single peer to define the behaviour of whole P2P system. System behaviour consists of sets of nodes, states, transitions, actions, and tasks. A node in a P2P network can have several states connected with a number of transitions in order to perform certain tasks and actions.

The characteristics of the nodes, transitions, actions, tasks and states are defined in the system scenario. The characteristics of nodes include the number of nodes used in the network, their connection details in terms of uplink downlink speeds, their behaviours in terms of loops, delays, and actions etc. Additional behaviours can also be added and used by calling the behaviour with CallBehaviour attribute. The resource allocation for each peer is also defined in the scenario. After a network has been setup, its corresponding code is generated in XML for both system behaviour and system scenario.

**Statistics**

PeerThing provides a GUI interface for its users to interact with the system. Simulation takes the system behaviour and scenario details as input and generates the output in a specified log file. PeerThing has the capability to run multiple simulations for a specified number of time steps or number of messages. The generated results can also be converted to .csv format and opened in Microsoft Excel.

**Scalability**

The authors have claimed that this simulator can simulate up to 2000 nodes for Gnutella network and 1000 nodes for Napster model. But the simulator can successfully run up to 700 nodes without generating any errors. A java heap space error is generated when more than 700 nodes are used to simulate a proper Gnutella network with its proper functionalities.

**Usability/ Documentation**

The user manual for PeerThing as compared to all the surveyed simulators can be considered as the best user manual. It has detailed information on how to use the simulator. It contains a step-by-step guide to build a basic network and view its corresponding results. The authors didn't provide any API documentation with the simulator source code. The source code is not well commented and too complex to understand as there are many things interconnected with each other. But the user manual contains sufficient information for a user to model a basic network.

**Portability**

The results generated from the simulator are always stored in a log file before another simulation is performed. The results generated in form of tables and their corresponding graphs in the simulator visualisation are limited to visualisation provided by the simulator. Such results are not sufficient enough for analysis on subjects not supported by the simulator. But the capability of exporting files in .csv format allows a much-detailed analysis to be done on the generated results.

**Extensibility**

The extensibility of the simulator code is an extremely hard task to do as the source code is not well commented. But extensibility can be performed in the behaviour of the provided Gnutella and Napster basic models. The default Gnutella behaviour provided in the simulator only performs searching mechanism and does not provide any functionality for download operations if the searched file is found.

*D. Results & Analysis*

The analyzing part can be performed based on the parameters of architecture & statistics, scalability, usability and extensibility.

The summary of overall simulator analysis is described in Table 1.

- Architecture defines that whether it has the capability to handle the discrete events for structured and unstructured networks.

It can be observed from the Table 2 that the architecture of Query Cycle, P2P & 3LS simulator are not well structured & developed, the accuracy and efficiency is not too good as well as compared to the other simulators, where as peer thing has the best So the best architecture to work is with Peer Thing.

- Furthermore it can be observed from Fig. 1 that the scalability can be best achieved in PeerSim and Query cycle simulator while moderate values were observed in Neuro and P2PSim simulator. It should be worthwhile to mention here that scalability is a secondary issue in considering the peer to peer networks as scalability refers to the number to nodes which can be accommodated , the primary concern is understand the architecture and ,check its extensibility and statistics against any desired network. Query Cycle and peer Sim can support million of nodes while PeerThing and RealPeer can support 2000 and 20,000 nodes respectively, the smallest of which is observed is with 3LS where it can only support 20 nodes.

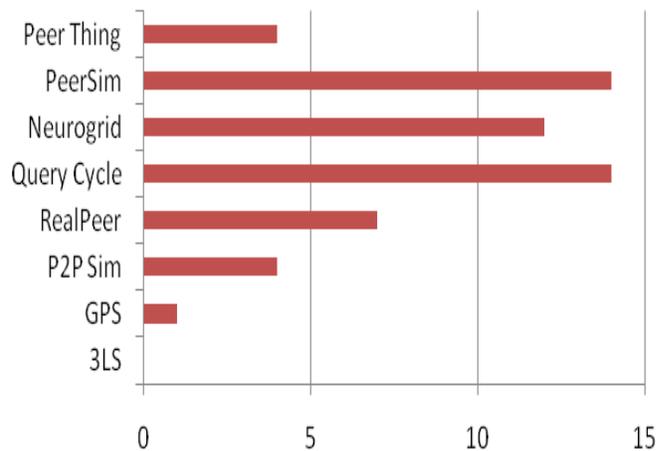

**Figure 1: comparison of simulators scalibility**

- Statistics in our paper refers to the GUI / non GUI environment and the status.

Fig. 2 shows that the pearSim and RealPeer are not equipped with the GUI environment while the rest of them are based on GUI environments. GUI provides a user friendly environment to perform the task else the only option which is left is to utilize the command line format which is not an easy task especially in complex networks.

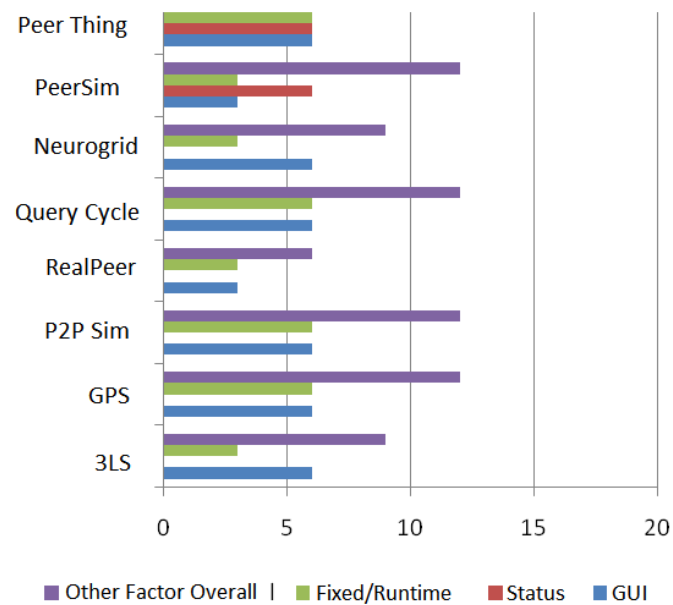

**Figure 2 : other factors comparison**

- Usability/ Documentation usually refer to the amount of data which is available in the form of material or tutorial to perform the desired task in a professional manner.

**Table 1: Summary of Simulator Analysis**

| Simulators | Architecture | Usability/ Documentation | Scalability | Extensibility | Statistics | Language |
|---|---|---|---|---|---|---|
| **P2Psim** | Discrete-event | Poor documentation | Upto 3000 nodes for Chord protocol | Limited protocol extensibility but complex to handle | GUI | C++ |
| **PeerSim** | Discrete-event | Detailed API documentation but lacks detail in user manual | Highly Scalable up to $10^6$ nodes supported | Designed to be extensible | No GUI | Java |
| **3LS** | Discrete-event | N/A | Low scalability of 20 nodes | Theoretically extensible but practical implementation is required | GUI | Java |
| **Neurogrid** | Discrete-event for unstructured networks | Good API documentation and user manual but source code is not commented well | Around 300,000 nodes are claimed | Designed to be extensible | GUI | Java |
| **GPS** | Discrete event for both structured and unstructured networks | Poor documentation | Up to 512 nodes tested | Lack of good documentation limits extensibility | GUI | Java |
| **Query Cycle** | | Poor API documentation and user manual | Highly scalable up to $10^6$ nodes supported | Limited | GUI | Java |
| **RealPeer** | Discrete Event | Well commented source code but poor API documentation and user manual | Around 20,000 peers can be simulated | Lack of good API documentation does not support extensibility | No GUI | Java |
| **PeerThing** | Discrete event driven for both unstructured and structured networks | Good User Manual but source code is not well commented and poor API documentation | Gnutella 2000 nodes & Napter 1000 | Extensible with and without changes made in source code | GUI | Java |

The graph in Fig. 3 shows that the best usability info are in Neuro and PeerThing while above moderate value are been recorded for PeerSim and , Real Peer simulators.

- Extensibility defines whether the simulator code can be modified according to the user requirement and how well the simulator work on available computer resources.
- Peerthing, Peersim and Neuro grid are highly extensible and efficiently make uses of the available resources where as RealPeer has the worst extensibility as shown in Fig. 4.

Table 2: Architecture Comparison

| Simulators | Structured/ Unstructured | Event/Cycle/Query based | Triggered by | Distributed Simulation | Dynamic Network | Special Features | Accuracy | Efficiency |
|---|---|---|---|---|---|---|---|---|
| **PeerSim** | Both | Event and Cycle based | Time and random selection | N/A | Yes | N/A | Moderate | Less |
| **GPS** | Both | Discrete Event based | Occurrence of Event | Yes | Yes | Macroscopic models | Highly | Highly |
| **Neurogrid** | Both | Discrete Event based | Time Scheduling | N/A | Yes | N/A | Moderate | Less |
| **Query Cycle** | Un Structured | Query Based | Time | N/A | Yes | For file sharing | Less | Less |
| **P2PSim** | Structured | Discrete Event based | Time | No | N/A | For understanding P2P Source code | Less | Less |
| **3LS** | Un Structured | Event based | Time Scheduling to user specific behavior | N/A | Yes | 3 Separate level Architecture; Network, Protocol and User | Less | Less |
| **RealPeer** | Un Structured | Discrete-event | Time Scheduling | N/A | N/A | Development environment; Reproducibility of results | Moderate | Moderate |
| **PeerThing** | N/A | Event based | Time Scheduling to user specific behavior | Yes | Yes | System behaviour and System Scenario Definition | Highly | Highly |

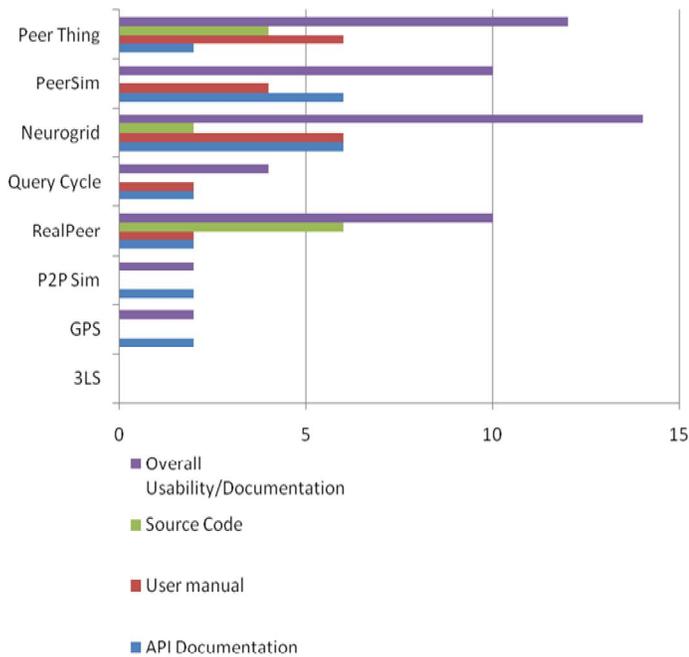

Figure 3 : Usability & documentation analysis

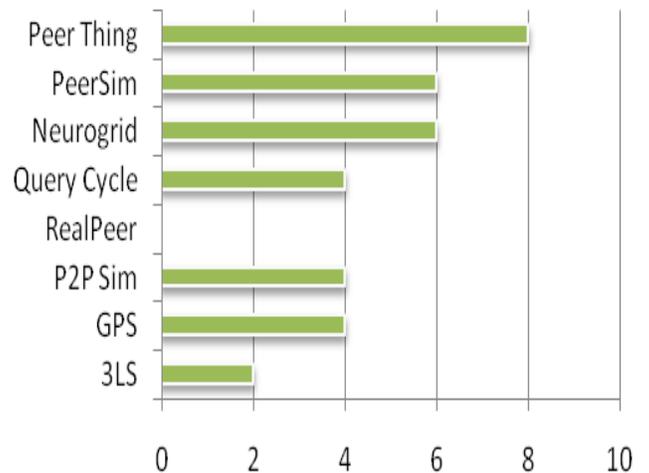

Figure 4: Extensibility comparison

Based on the discussion provided above and by observing the overall graph in Fig. 5, we can conclude by mentioning that the best simulated results are observed in Peer-Thing simulator and furthermore it provides more user friendly environment.

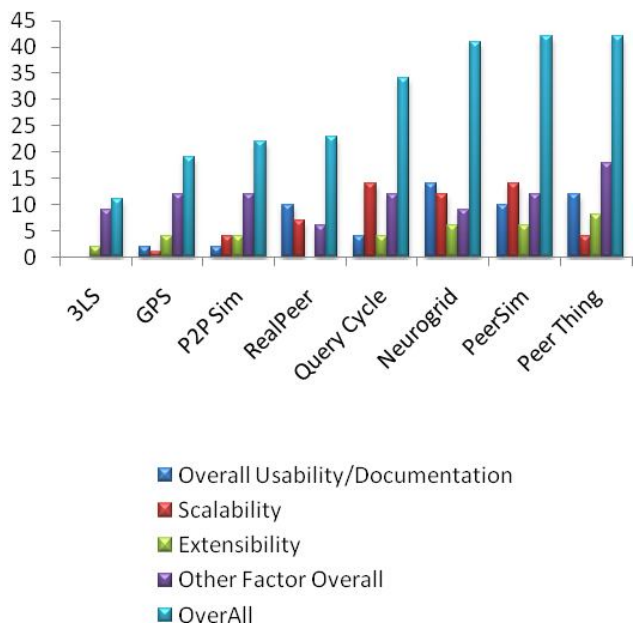

**Figure 5: overall parameters comparision of simulators**

Some detail information regarding the PeerThings is mentioned below.

PeerThing [14] simulator is an application for modelling and simulating peer-to-peer-networks. PeerThing is written on Java and a tool known as Eclipse is used to provide a GUI interface. One of the most important feature of PeerThing is that it is flexible i.e. it allows the modelling of different architectures. Results generated from PeerThing simulations can be stored in a database and can be viewed by manipulating SQL-statements as well as it also has the capability of comparing different simulations, even new P2P systems can also be designed with this application [15].

PeerThing is very easy to handle and use, as there is no need to install the PeerThing simulator as the PeerThing developer has made an executable file of the simulator, execute the .exe file and the simulator starts. PeerThing is a graphical based simulator that made it easy to use. The user has to perform three basic steps to perform any simulation. The simulator contains a detailed user manual to assist its users to perform simulations. Its online support is also available through group mails since it is currently in active state. Implementation of new scenarios does not require changes in the source code of the simulator; therefore it possesses the capability of modelling virus behaviour in P2P networks.

## CONCLUSION

This paper gives a new course in selecting P2P simulators on analysis basis. The selected simulator was better than other P2P simulators in terms of documentation, statistics usability, extensibility and Portability, however there were few weakness in which the main was regarding scalability as it limits the number of peer in the network and if the peers exceeds the limit the performance of the simulator becomes worse and at certain point it results in error. As the number of peers in the simulator is increased than the simulation time steps also starts decreasing at this point the desired simulations results are hard to obtain, as the simulator is not able to perform the desired amount of simulation to analysis the performance. Further these problems can be over come in future by analysis the above discussed parameters.

ACKNOWLEDGMENT

The authors would like to acknowledge the support of Queen Mary, University of London and Iqra University.